\documentclass[12pt]{article}
\usepackage[a4paper, total={7in, 10in}]{geometry}
\usepackage[parfill]{parskip}
\usepackage{physics, tensor, float, subcaption}
\usepackage{graphicx}
\graphicspath{ {Plots/} }
\usepackage{jhep-mod}
\usepackage{bm}
\usepackage{soul}
\usepackage{amssymb,amsmath,amsthm}
\usepackage{mathrsfs}
\usepackage[utf8]{inputenc}
\usepackage{enumerate}
\usepackage{bigints}
\usepackage{xcolor}
\usepackage{appendix}
\usepackage{graphicx}
\usepackage{float}
\usepackage{tikz}
\usepackage{setspace}
\usepackage{cancel}
\usepackage{array}
\usepackage{tabulary}
\usepackage{doi}
\usepackage{comment} %to have sections commented out======
\definecolor{purple}{rgb}{1,0,1}
\definecolor{lime}{HTML}{A6CE39} % needs xcolor
\definecolor{lime}{HTML}{A6CE39}
\newcommand{\orcidicon}{%
	\begin{tikzpicture}
	\draw[lime, fill=lime] (0,0) 
		circle [radius=0.16] 
		node[white] {{\fontfamily{qag}\selectfont \tiny ID}};
	\draw[white, fill=white] (-0.0625,0.095) 
		circle [radius=0.007];
	\end{tikzpicture}
	\hspace{-5mm}
}
\newcommand\orcidRudeep{{\href{https://orcid.org/0009-0002-0162-562X}{\orcidicon}}}
\newcommand\orcidMatt{{\href{https://orcid.org/0000-0003-1088-6485}{\orcidicon}}}
\renewcommand{\O}{\mathcal{O}}

\newcommand{\be}{\begin{equation}}
\newcommand{\ee}{\end{equation}}

\protected\def\verythinspace{%
  \ifmmode
    \mskip0.5\thinmuskip
  \else
    \ifhmode
      \kern0.08334em
    \fi
  \fi
}

\newcommand{\thin}{\verythinspace}

\newcommand{\D}{\mathrm{d}}

\newcommand{\riem}[2]{R^{\,#1}{_{#2}}}

\newcommand{\weyl}[2]{C^{\,#1}{_{#2}}}
\newcommand{\laund}[1]{\mathcal{O}{\left(#1\right)}}
\newcommand{\ds}[2]{g_{#1#2}\,\D x^{#1}\D x^{#2}}
\newcommand{\metric}[3]{g^{#1}{_{#2#3}}}
\newcommand{\metcomp}[1]{{{\Delta}}_{#1}}

\begin{document}
%========================================================

%========================================================
%========================================================

\title{\vspace{-25pt}\huge{
Black holes embedded in FLRW cosmologies\\
}}

%========================================================
%========================================================

%========================================================
\author{
\Large
Rudeep Gaur\!\orcidRudeep {\sf  and} Matt Visser\!\orcidMatt}
%========================================================
%========================================================
%========================================================
%========================================================
\affiliation{School of Mathematics and Statistics, Victoria University of Wellington, 
\\
\null\qquad PO Box 600, Wellington 6140, New Zealand.}
%========================================================
%========================================================
\emailAdd{rudeep.gaur@sms.vuw.ac.nz}
\emailAdd{matt.visser@sms.vuw.ac.nz}
%========================================================
%========================================================

\abstract{
\vspace{1em}

There has recently been some considerable interest expressed in a highly speculative model of black hole evolution --- allegedly by a postulated direct coupling between black holes and  cosmological expansion \emph{independently of accretion or mergers}. We wish to make several cautionary comments in this regard --- at least three exact solutions corresponding to black holes embedded in  a FLRW background are known, (Kottler, McVittie, Kerr-de~Sitter), and they show no hint of this claimed effect --- thereby implying that this claimed effect (if it exists at all) is certainly nowhere near  ubiquitous. 

\bigskip
\noindent
{\sc Date:} 15 August 2023; 21 August 2023; \LaTeX-ed \today

\bigskip
\noindent{\sc Keywords}:\\
Black holes; FLRW cosmologies; Schwarzschild-de~Sitter (Kottler);  \\
Schwarzschild-FLRW (McVittie);  Kerr-de~Sitter.

\bigskip
\noindent{\sc PhySH:} 
Gravitation
}

%========================================================
\maketitle
%========================================================
\def\tr{{\mathrm{tr}}}
\def\diag{{\mathrm{diag}}}
\def\cof{{\mathrm{cof}}}
\def\pdet{{\mathrm{pdet}}}
\def\QED{ {\hfill$\Box$\hspace{-25pt}  }}
\def\d{{\mathrm{d}}}
\parindent0pt
\parskip7pt
%=====================================================
\clearpage
%=====================================================
\section{Introduction}
%=====================================================

There have recently been some rather bold and unusual claims made regarding how black holes  might  \emph{directly} interact with the overall FLRW cosmological expansion~\cite{Farrah:2023}. (See also the somewhat earlier closely related references~\cite{Croker:2019a,Croker:2019b,Croker:2020} which developed the theoretical framework for these claims.) 

Key parts of the claims made in reference~\cite{Farrah:2023} were that:
\begin{itemize}
\item ``The Kerr black hole solution is ... \emph{provisional} as its behavior at infinity is incompatible with an expanding universe.'' [emphasis added].
\item
``Black hole models with realistic behavior at infinity predict that the gravitating mass of a black hole can increase with the expansion of the universe \emph{independently of accretion or mergers}...''  [emphasis added].
\item ``The redshift dependence of the mass growth implies that, at $z\leq7$, \emph{black holes contribute an effectively constant cosmological energy density} to Friedmann's equations.'' [emphasis added].
\end{itemize}
There are a number of significant problems with these claims:
\begin{itemize}
\item 
The truly enormous ``separation of scales'' that is observed to occur between galactic dynamics and cosmological dynamics makes all such claims grossly implausible. (More on this specific point below.)
\item
There are at least three exact solutions to the Einstein equations that embed black holes in expanding universes, (Kottler, McVittie, and Kerr--de~Sitter), and in those known exact solutions the claimed effect simply does not occur. (This will be the main point of the current article.)
\item
The underlying theoretical framework~\cite{Croker:2019a,Croker:2019b,Croker:2020} adopted in reference~\cite{Farrah:2023} appears to be deeply flawed~\cite{Mistele:2023}. 
(One key issue here is that the cosmological mass fraction sequestered in black holes simply does \emph{not} lead to an equal but opposite pressure; a ``black hole gas'' mimics ``dust'', it does not mimic ``dark energy''.)  Several other authors have made related cautionary comments~\cite{Parnovsky,Avelino}.
\item
An independent observational analysis~\cite{LeiLei:2023} strongly excludes the claimed effect at $\sim 3\sigma$, and is compatible with zero  effect at $\sim 1\sigma$. (The technical difficulty with making this bound even tighter  lies in guaranteeing that the observational sample is free of false positives, due to the possible growth of superficially quiescent black holes actually being driven due to some unaccounted for variant of the usual processes of \emph{accretion} and/or \emph{mergers}.)
Several other independent observational and/or numerical analyses similarly disfavour the existence of the claimed effect, see for instance~\cite{Rodriguez:2023,Andrae:2023,Amendola:2023, Easther:2023, Clifton:2018}. 
\end{itemize}

\clearpage
In the current article we will concentrate on the general relativistic aspects of the situation, paying particular attention both to physically relevant approximations, and to the known exact theoretical solutions --- we will argue that based on the known exact solutions there is simply no physical reason to expect the claimed effect to occur, and good physics reasons to reject the claimed effect.

%============================================
\section{Separation of scales}\label{scale_section}
%============================================
%%% 1 parsec = 3 x 10^{16} m = 3 x 10^{13} km
%%% (giga solar mass BH) ~ 3 x 10^9 km ~ 10^{-4} parsec.
%============================================

We start the discussion by pointing out that there is a truly enormous separation of scales between galactic black hole physics and cosmological physics. 
Even the heaviest known galactic black holes have masses only of order $3\times 10^{10}$ solar masses, corresponding to a Schwarzschild radius $\lesssim 10^{-3}$ parsec. In contrast the cosmological homogeneity scale is typically taken to be of order $\gtrsim 10^{8}$ parsecs\footnote{This is usually called the \textit{statistical scale of homogeneity} (SSH) and estimates thereof are  most often based on the  galaxy-galaxy   2--point correlation method.}, and the Hubble scale is even larger, of order $10^{10}$ parsec. There simply is no plausible mechanism for directly coupling milli-parsec black hole physics to giga-parsec cosmological physics. 
(For related comments see references~\cite{Sadeghi:2023,Wang:2023}.)

What is much more plausible is to directly couple the observed black hole candidates found in most spiral galaxy cores  to matter in their immediate environment --- the galactic cores and  galaxies in which they reside. But this of course implies black hole evolution due to the utterly standard processes of \emph{accretion} and/or \emph{ mergers}, which is exactly what the authors of reference~\cite{Farrah:2023} are claiming to side-step.

More quantitatively, even in the absence of an explicit exact solution to the Einstein equations, we can argue as follows: Any attempt at inserting a black hole into a FLRW cosmology will at the very least involve two separate mass scales --- $m$ the mass of the central black hole, and $\rho_{\scriptscriptstyle{\mathrm{FLRW}}}\, r^3$, the FLRW contribution to the mass contained in a ball of radius $r$. Combining these two quantities  defines a natural distance scale
\begin{equation}
r_* = \sqrt[3]{m\over \rho_{\scriptscriptstyle{\mathrm{FLRW}}} }.
\end{equation}
At distances $r<r_*$ black hole physics dominates, at distances $r>r_*$ the FLRW cosmology dominates. 
We shall see this natural transition-distance scale crop up repeatedly in the discussion below.

\clearpage
%========================================
\section{Relevant exact solutions in general relativity}
%========================================

There are at least three well-known \emph{exact} solutions of Einstein's equations for black holes embedded in expanding FLRW universes:
\begin{itemize}
\item Schwarzschild--de~Sitter (Kottler);
\item Schwarzschild--FLRW (McVittie);
\item Kerr--de~Sitter.
\end{itemize}
Note that de Sitter spacetime is just a special case of FLRW,  which, \emph{in appropriate coordinates}, corresponds to exponential expansion $a(t) = e^{Ht}$, with constant Hubble parameter $H$. Furthermore, since in the standard framework of $\Lambda$CDM cosmology, the universe at the current epoch is believed to already be cosmological constant dominated, it follows that de~Sitter space is an excellent approximation to both the near-current-epoch  and future expansion history of the universe. Allowing for a completely  arbitrary expansion history for the scale factor $a(t)$,  [as in the Schwarzschild--FLRW (McVittie) spacetime discussed below], while it would be ``nice to have'', is not really critical for purposes of the current discussion.

In the discussion below the only even slightly tricky part is to keep careful track of a number of suitable coordinate transformations. A useful recent reference explicitly discussing coordinate freedom in cosmological spacetimes is~\cite{PG-cosmology}. 

%===============================================
\subsection{Schwarzschild--de~Sitter (Kottler)}
%===============================================
\label{SS:Kottler}
%==================================================

Let us start from Schwarzschild--de~Sitter (Kottler)  spacetime presented in its most common form, in static $(t,r)$ coordinates~\cite{Kottler}:
\begin{equation}
\d s^2 = -\left(1-{2m\over r}-H^2 r^2\right) \d t^2 + {\d r^2\over 1-{2m\over r}-H^2r^2 } + r^2 \d\Omega^2.
\end{equation}
This coordinate system makes it obvious that at small $r$ one recovers the Schwarzschild solution, and that the mass $m$ of the central black hole is not changing.
In Kottler spacetime   the natural distance scale (in physical units) reduces to
\begin{equation}
r_* = \sqrt[3]{m\over \rho_{\scriptscriptstyle{\mathrm{FLRW}}} } \;
\longrightarrow\; \sqrt[3]{m c^2\over H^2 },
\end{equation}
and can be identified as the radius of the OSCO, (the \emph{outermost stable circular orbit})~\cite{Boonserm:2019,Berry:2020}. For $r\ll r_*$ the physics is black-hole dominated, for $r \gg r_*$ the physics is cosmological-constant dominated.

%========
\clearpage
Now this particular static slicing makes the physical interpretation in terms of an exponentially expanding FLRW spacetime not entirely obvious --- we need to make a few coordinate transformations to make this fully explicit. 
To proceed with the discussion, we first substitute
\begin{equation}
\d t = \d\bar t + {\sqrt{ {2m\over r}+H^2 r^2} \over 1-{2m\over r}-H^2 r^2} \;\d r,
\end{equation}
to go to $(\bar t, r)$ Painlev\'e--Gullstrand coordinates (see for instance reference~\cite{PG-cosmology}):
\begin{equation}
\d s^2=-\d\bar t^2 + \left[ \d r - \sqrt{ {2m\over r}+H^2 r^2}\; \d\bar t\right]^2 +r^2 \d\Omega^2.
\end{equation}
Second, we now set $r=e^{H\bar t}\; \bar r$, so that $\d r = e^{H\bar t}\; [ \d\bar r + \bar r H \d\bar t]$. Then in these new $(\bar t, \bar r)$ coordinates
\begin{equation} 
\d s^2 = -\d\bar t^2 + e^{2H\bar t} \left[ 
\left( \d\bar r + \left\{ H \bar r - \sqrt{ {2m\;e^{-3H\bar t}\over \bar r}+H^2 \bar r^2} \right\}\d\bar t\right) ^2 + \bar r^2 \d\Omega^2 \right].
\end{equation}
Note that at large distances
\begin{eqnarray}
\left\{ H \bar r - \sqrt{ {2me^{-3H\bar t}\over \bar r}+H^2 \bar r^2} \right\} 
&=& 
H \bar r \left\{ 1 - \sqrt{ 1 +{2me^{-3H\bar t}\over H^2 \bar r^3}} \right\}
\nonumber\\ 
&=&
-{m\over H\bar r^2}\; e^{-3Ht} 
+ \O\left(1\over \bar r^{5}\right).
\end{eqnarray}
Thus at large $\bar r$
\begin{equation} 
\d s^2 = -\d\bar t^2 + e^{2H\bar t} \left[ 
\d\bar r ^2 + \bar r^2 \d\Omega^2 \right] + \O\left(1\over\bar r^2\right),
\end{equation}
making it obvious that the Schwarzschild--de~Sitter (Kottler) black hole is embedded in an exponentially expanding FLRW universe. 

We emphasize that in this specific example there simply is no coupling between the mass parameter $m$ and the cosmological parameter $H$; they are independent constants.  
The only even slightly tricky part of the analysis was in setting up the coordinate transformations used to make these properties manifest. 

\clearpage 
\null
\vspace{-75pt}
%===========================================
\subsection{Schwarzschild--FLRW (McVittie)}
%============================================
\label{SS:McVittie}
%============================================

The Schwarzschild--FLRW (McVittie) spacetime metric~\cite{McVittie,Kaloper:2010,Lake:2011,Faraoni:2012} describes an eternal black hole that has been part of the universe ever since the Big Bang --- if in contrast one wants to describe a black hole that forms from stellar collapse, then a \emph{segment} of the Schwarzschild--FLRW (McVittie) spacetime should be used --- only for describing the quiescent period after the initial collapse and ringdown.

McVittie spacetime can be represented in any of the following four completely equivalent forms~\cite{PG-cosmology}:
\begin{equation}
\label{E:McVittie-standardA}
\d s^2 = - \left(1-{m\over 2 a(t) \tilde r} \over 1+{m\over 2 a(t) \tilde r} \right)^2 \d t^2 
+ \left(1+{m\over 2 a(t) \tilde r} \right)^4 a(t)^2 
\{ \d \tilde r^2 + \tilde r^2 \d\Omega^2\}.
\end{equation}
\begin{equation}
\label{E:McVittie-CPGA}
\d s^2 =   \left(1+{m\over 2 \bar r} \right)^4   \left\{ - \left([1-{m\over 2 \bar r}]^2 \over [1+{m\over 2 \bar r}]^6 \right) \d t^2 
+
\{ [\d \bar r- H(t)\bar r\d t]^2 + \bar r^2 \d\Omega^2\}\right\} .
\end{equation}
\begin{equation}
\label{E:McVittie-KKM-bA}
\d s^2 =
- \left(1-{2m\over \hat r}\right)  \d t^2 
+ \left[{\d \hat r\over\sqrt{1-2m/\hat r}} -  H(t) \hat r \d t\right]^2
+ \hat r^2 \d\Omega^2.
\end{equation}
\begin{equation}
\label{E:McVittie-comovingA}
\d s^2 =
- \left(1-{2m\over a(t) r}\right) \d t^2 
+ a(t)^2 \left\{ \left[{\left(\d r + H(t) r \left[1-\sqrt{1-{2m\over a(t) r}}\right]  \d t\right)^2\over{1-{2m\over a(t) r}}} \right]
+  r^2 \d\Omega^2 \right\}.
\end{equation}
All four of these coordinate systems use the same time coordinate $t$, and also the same angular coordinates $\{\theta,\phi\}$, while we have used coordinate freedom of general relativity to adopt differing radial coordinates $\{\tilde r,\bar r, \hat r, r\}$.
(The relevant coordinate transformations connecting these differing radial coordinates are explicitly presented in reference~\cite{PG-cosmology}.)

In all four of these coordinate systems the energy density is determined by finding the time-like eigenvector of the stress-energy, and is easily calculated to be~\cite{PG-cosmology}:
\begin{equation}
\rho= {3\over8\pi} H(t)^2.
\end{equation}
The pressure is determined by the space-like eigenvectors of the stress energy and is more subtle:
Depending on which of the coordinate systems (\ref{E:McVittie-standardA})--(\ref{E:McVittie-comovingA}) one adopts one finds the superficially differing but physically equivalent results~\cite{PG-cosmology}: \enlargethispage{50pt}
\begin{eqnarray}
p 
&=& -\rho - {1\over4\pi} \; {1+{m\over 2a(t)\tilde r}\over 1-{m\over 2a(t)\tilde r}} \; {\dot H(t)} ;
\\
&=& -\rho- {1\over 4\pi} \; {1+{m\over 2\bar  r}\over 1-{m\over 2\bar r}} \;\dot H(t);
\\
&=&-\rho -  {1\over 4\pi} \;{1\over\sqrt{1-2m/\hat r}} \;\dot H(t);
\\
&=& -\rho - {1\over 4\pi} \;{1\over\sqrt{1-{2m\over a(t)r}}}\; \dot H(t).
\end{eqnarray}

\clearpage
At large distances, in all four of these coordinate systems, one recovers the standard spatially flat ($k=0$) FLRW result:\\
\begin{equation}
p \to -\rho - {\dot H(t)\over4\pi}.
\end{equation}

Turning now to the explicit representations of the  spacetime metric, at suitably large distances, $a(t)\tilde r \gg m$, the line element (\ref{E:McVittie-standardA}) implies
\begin{equation}
\label{E:McVittie-standardA2}
\d s^2 \approx - \d t^2 + a(t)^2  \{ \d \tilde r^2 + \tilde r^2 \d\Omega^2\},
\end{equation}
which clearly is ($k=0$) FLRW with arbitrary scale factor $a(t)$. 

On the other hand at suitably small distances, $\bar r H(t) \ll 1$,  the line element (\ref{E:McVittie-CPGA}) implies
\begin{equation}
\label{E:McVittie-CPGA2}
\d s^2 \approx   \left(1+{m\over 2 \bar r} \right)^4   
\left\{ - \left([1-{m\over 2 \bar r}]^2 \over [1+{m\over 2 \bar r}]^6 \right) \d t^2 
+
\{ \d \bar r^2 + \bar r^2 \d\Omega^2\}\right\} .
\end{equation}
This is just Schwarzschild spacetime in isotropic coordinates. 
So the mass of the central black hole is simply $m$, a time-independent constant.
Note there is no mass flux onto the central black hole; there is no accretion.
That is, there simply is no direct coupling between the mass parameter $m$ and the cosmological parameter $H(t)=\dot a(t)/a(t)$; they are independent quantities.  
As previously noted, the only even slightly tricky part of the analysis was in setting up the coordinate transformations used to make these properties manifest.

%====================================
\subsection{Kerr--de~Sitter}
%====================================

Rotating black holes are much more subtle than their non-rotating counterparts. The basic asymptotically flat Kerr spacetime was first discovered some 60 years ago in 1963, see reference~\cite{Kerr:1963}. Further discussion can be found in~\cite{Kerr:1965} and~\cite{Newman:1965,Newman-Janis, Boyer:1966, Carter:1968, Bardeen:1970}, and more recently in references~\cite{Kerr:book, Kerr:intro, Kerr:2007, Rajan:2016, Kerr:ansatz, Baines:3-function,Kerr:review1, Kerr:review2}.

The Kerr--de Sitter geometry is even more subtle than Kerr, and was first obtained by Carter some 10 years later in 1973;
 still some 50 years ago, see references~\cite{Carter:KdS,Carter:KdS2}. 
The Kerr--de~Sitter geometry represents an eternal rotating black hole embedded in de~Sitter spacetime.
For a recent easily accessible discussions see reference~\cite{Matzner}, and even more recently see~\cite{Li:2020,Hintz:2016}.

For a black hole formed from stellar collapse, one should certainly wait until after the initial collapse and ringdown, until the black hole is quiescent, and also wait until the universe is old enough to be cosmological constant dominated --- as is now expected to be the situation in the current epoch. That is, the Kerr--de Sitter geometry should be a good approximation to rotating black holes in the current epoch. (This point is implicit in the discussion of reference~\cite{Matzner}.) 

\clearpage
The metric for the Kerr--de~Sitter spacetime is most typically presented in stationary coordinates~\cite{Carter:KdS,Carter:KdS2}: 
%=======================================
%=================================
%=====================================================================
\begin{equation}
    \begin{aligned}
        ds^2 = &- {(r^2 + a^2)(1-\frac{\Lambda}{3}r^2) - 2mr\over r^2+a^2\cos^2\theta} 
        \left[ \D t - a\sin^2\theta\,\D\phi\over 1+{1\over3} \Lambda a^2\right]^2\\
        &\hspace{5em}+\sin^2\theta\left[1+{1\over3} \Lambda a^2\cos^2\theta\over r^2+a^2\cos^2\theta\right]
        \left[ a \D t - (r^2+a^2) \D\phi\over1+{1\over3}\Lambda a^2\right]^2 \\
        &\hspace{1em}+ (r^2+a^2\cos^2\theta) \left[ {\D r^2\over{(r^2 + a^2)(1-\frac{\Lambda}{3}r^2) -2mr}}
        + {\D \theta^2\over(1+{1\over3} \Lambda a^2 \cos^2\theta)}\right].
    \end{aligned}
\end{equation}
(Warning: Here $a$ is the spin parameter $a=J/m$, not the FLRW scale factor $a(t)$.)
In this spacetime, the cosmological constant is related to the Hubble \emph{constant} by $\Lambda = 3 H^2$. A perhaps mildly surprising aspect of this line element is the presence of the constant $1+{1\over3} \Lambda a^2 = 1+ H^2 a^2$ in several strategic places.

To be able to efficiently use computer algebra packages, it is more beneficial to have this metric in a fully expanded form, and to eliminate the trigonometric functions. We therefore re-write the line element in the following form: 
\begin{equation}\label{used kerr form}
    \begin{aligned}
        \ds{\mu}{\nu} = &- \Bigg[\frac{\metcomp{r} - \metcomp{\theta}\,a^2 (1-\chi^2)}{\rho^2\,\Xi^2 }\Bigg]\,\D t^2 + \frac{\rho^2}{\metcomp{r}}\,\D r^2 + \frac{\rho^2}{\metcomp{\theta}\thin (1-\chi^2)}\,\D \chi^2\\
        & + \frac{(1-\chi^2)}{\rho^2\,\Xi^2}\Big[\metcomp{\theta}\thin(r^2 + a^2)^2 - \metcomp{r}\thin a^2\thin (1-\chi^2)\Big]\, \D \phi^2\\
        &- \frac{2a(1-\chi^2)}{\rho^2\,\Xi^2} \Big[ \metcomp{\theta}\thin (r^2 + a^2) - \metcomp{r}\Big]\,\D t \D\phi\,.\\
    \end{aligned}
\end{equation}
Here 
\begin{equation}\label{metric comps}
    \begin{aligned}
        &\chi = \cos\theta;\\
        &\metcomp{r} = r^2 + a^2 - 2mr + \frac{\Lambda}{3}r^2\thin(r^2 + a^2);\\
        &\metcomp{\theta} = 1 + \frac{\Lambda}{3}a^2\cos^2\theta = 1 + \frac{\Lambda}{3}a^2\chi^2;\\
        &\rho^2 = r^2 +a^2\thin\cos^2\theta = r^2 + a^2\chi^2;\\
        &\Xi = 1 +\frac{\Lambda}{3}a^2\,.\\
    \end{aligned}
\end{equation}

The Kerr-de Sitter spacetime is a $\Lambda$-vacuum solution of the Einstein field equations, an Einstein manifold, and therefore satisfies \enlargethispage{20pt}
\begin{equation}\label{Einstein tensor check}
    R_{\mu\nu} = -\Lambda\metric{}{\mu}{\nu}; \qquad 
    G_{\mu\nu} = +\Lambda\metric{}{\mu}{\nu} .
\end{equation}
Using, for example, \texttt{sagemath} or \texttt{Maple}, we may easily check this is in fact true. 

\clearpage

We must also check that the Weyl tensor is nonzero, and that the Weyl scalar, $\weyl{\mu\nu\alpha\beta}{}\weyl{}{\mu\nu\alpha\beta}$ is position-dependent: Indeed
\begin{equation}
     \weyl{\mu\nu\alpha\beta}{}\weyl{}{\mu\nu\alpha\beta}= -\frac{m^2 (a^2\chi^2 + 4ar\chi + r^2)\,(a^2\chi^2 - 4ar\chi+ r^2)\,(r^2-a^2\chi^2)}{(r^2+a^2\chi^2)^{6}},
\end{equation}
which depends on both $r$ and $\chi$.
Furthermore, the Kretschmann scalar, $\riem{\mu\nu\alpha\beta}{}\riem{}{\mu\nu\alpha\beta}$ is also non-zero and position-dependent. 
Lastly, due to the Kerr--de Sitter spacetime no longer being a pure vacuum solution to the Einstein equations, we expect the difference between the Kretschmann scalar and Weyl scalar to be non-zero (and position-independent). We find
\begin{equation}
    \riem{\mu\nu\alpha\beta}{}\riem{}{\mu\nu\alpha\beta} - \weyl{\mu\nu\alpha\beta}{}\weyl{}{\mu\nu\alpha\beta} = \frac{8}{3}\,\Lambda^2\,.
\end{equation}\enlargethispage{15pt}
When $\Lambda = 0$ --- where the Kerr metric is recovered ---  the difference is zero. 

We will subsequently look at the asymptotic large-distance behaviour and verify that in a suitable coordinate system the cosmological constant $\Lambda$ can be reinterpreted in terms of a constant Hubble parameter $H$ (with $\Lambda = 3H^2$), and an exponentially growing scale factor $a(t)=e^{Ht}$. 
But for now let us focus on a number of internal consistency checks for the Kerr--de~Sitter spacetime.

%===========================================================
\section{Extended consistency checks for Kerr--de~Sitter}
%===========================================================
\label{consistency checks}
%===========================================================

In this section we shall spend a little effort checking that the Kerr-de Sitter spacetime does in fact (under suitable circumstances) reduce to the Kerr, Kottler, and de~Sitter spacetimes as required. 

%=============================
\subsection{Kerr spacetime}
%=============================
We first investigate the $\Lambda=0$ limit of the KdS metric given in \eqref{used kerr form}, resulting in the Kerr spacetime. This results in a vacuum solution to the Einstein equations, hence, providing the basis for a variety of consistency checks for the KdS spacetime. 

When $\Lambda = 0$ we obtain the line element
\begin{equation}\label{lambda zero}
    \begin{aligned}
        \ds{\mu}{\nu} = &- \Bigg[\frac{\bar{\Delta}_r - \,a^2 (1-\chi^2)}{\rho^2 }\Bigg]\,\D t^2 + \frac{\rho^2}{\bar{\Delta}_r}\,\D r^2 + \frac{\rho^2}{(1-\chi^2)}\,\D \chi^2\\
        &+ \frac{(1-\chi^2)}{\rho^2}\Big[\thin(r^2 + a^2)^2 - \bar{\Delta}_r\thin a^2\thin (1-\chi^2)\Big]\, \D \phi^2\\
        &- \frac{2a(1-\chi^2)}{\rho^2} \Big[  (r^2 + a^2) - \bar{\Delta}_r\Big]\,\D t \D\phi\,,\\
    \end{aligned}
\end{equation}
where $\bar{\Delta}_r = r^2 + a^2 - 2mr$.

To check this is in fact the Kerr spacetime, we compute the curvature quantities such as the Ricci tensor, Weyl tensor, Kretschmann scalar, and Weyl scalar. For a vacuum solution we expect the Ricci tensor to be zero and, therefore, the Riemann tensor and Weyl tensor to be equal. We verify this using \texttt{sagemath}/\texttt{Maple}. Furthermore, we find that the Riemann tensor and Weyl tensor are equal, non zero and position dependent. Lastly, the Weyl scalar and Kretschmann scalar are equal (as expected).

%====================================
\subsection{Kottler Spacetime}
%====================================
In the $a \to 0$ limit of the KdS spacetime, we recover the Kottler (Schwarzschild--de Sitter) spacetime. Firstly
\begin{equation}
\metcomp{r} \to \hat{\Delta}_r = r^2-2mr - {\Lambda r^4\over3} 
= r^2\left(1-{2m\over r} - {1\over3} \Lambda r^2 \right).
\end{equation}

Then 
\begin{equation}
(\D s^2)_{Kottler} = - {\hat{\Delta}_r\over r^2} 
\left[ \D t \right]^2
+{(1-\chi^2) r^2} 
\left[ \D\phi\right]^2 
+ r^2 \left[ {\D r^2\over\hat{\Delta}_r }
+ {\D\chi^2\over (1-\chi^2)}
\right]. 
\end{equation}
Rewritten, this becomes
\begin{equation}
(\D s^2)_{Kottler} = 
- {\hat{\Delta}_r\over r^2} \d t^2 + {r^2\over\hat{\Delta}_r} \D r^2
+ r^2 \left[ {\D \chi^2 \over(1-\chi^2)} + (1-\chi^2) \D\phi^2\right]\,.
\end{equation}
That is,
\begin{eqnarray}
(\D s^2)_{Kottler} &=&
- \left(1-{2m\over r} - {1\over3} \Lambda r^2 \right) \D t^2 
+ {\D r^2\over 1-{2m\over r} - {1\over3} \Lambda r^2} 
\nonumber\\
&&\qquad
+ r^2 \left[ \frac{\D\chi^2 }{(1-\chi^2)} + (1-\chi^2) \D\phi^2\right].
\end{eqnarray}
This metric is evidently the Kottler spacetime \cite{Kottler} in standard $(t,r)$ coordinates, (and not entirely standard $(\chi,\phi)$ coordinates). We may now perform the same consistency checks on this spacetime as we did in the KdS case. We expect similar results, as it is no longer a pure vacuum solution and corresponds to pure cosmological constant. We again find 
\begin{equation}
    R_{\mu\nu} = \Lambda \,g_{\mu\nu}\,; \qquad
    G_{\mu\nu} = -\Lambda \, g_{\mu\nu}\,.
\end{equation}
The curvature quantities such as the Riemann tensor and Weyl tensor are not equal, they are again non-zero and position-dependent.

The Kretschmann scalar is
\begin{equation}\label{kottler Kretschmann}
    \riem{\mu\nu\alpha\beta}{}\riem{}{\mu\nu\alpha\beta} = \frac{8}{3}\Lambda^2 + \frac{48m^2}{r^6}\,,
\end{equation}
and the Weyl scalar is
\begin{equation}\label{kottler weyl}
    \weyl{\mu\nu\alpha\beta}{}\weyl{}{\mu\nu\alpha\beta} = \frac{48m^2}{r^6}\,.
\end{equation}
The difference is simply
\begin{equation}
    \riem{\mu\nu\alpha\beta}{}\riem{}{\mu\nu\alpha\beta} - \weyl{\mu\nu\alpha\beta}{}\weyl{}{\mu\nu\alpha\beta} = \frac{8}{3}\,\Lambda^2\,,
\end{equation}
which is the same result as in the KdS case, which is to be expected since in the KdS case, the difference did not depend on the angular momentum. 

%=====================================
\subsection{de Sitter Spacetime}
%=====================================
The last parameter we shall set to zero is the mass of the black hole, $m \to 0$. The only change in the metric components is that now
\begin{equation}
 \metcomp{r}\to \Tilde{\Delta}_r = r^2 + a^2 - \frac{\Lambda r^2}{3}(r^2 +a^2) = (r^2 + a^2)\,\Big(1 - \frac{1}{3}\Lambda a^2\Big)\,.
\end{equation}

The KdS line element now reduces to
\begin{equation}\label{m=0 form}
    \begin{aligned}
        \ds{\mu}{\nu} = &- \Bigg[\frac{\Tilde{\Delta}_r - \metcomp{\theta}\,a^2 (1-\chi^2)}{\rho^2\,\Xi^2 }\Bigg]\,\D t^2 + \frac{\rho^2}{\Tilde{\Delta}_r}\,\D r^2 + \frac{\rho^2}{\metcomp{\theta}\thin (1-\chi^2)}\,\D \chi^2\\
        &+ \frac{(1-\chi^2)}{\rho^2\,\Xi^2}\Big[\metcomp{\theta}\thin(r^2 + a^2)^2 - \Tilde{\Delta}_r\thin a^2\thin (1-\chi^2)\Big]\, \D \phi^2\\
        &- \frac{2a(1-\chi^2)}{\rho^2\,\Xi^2} \Big[ \metcomp{\theta}\thin (r^2 + a^2) - \Tilde{\Delta}_r\Big]\,\D t \D\phi\,.\\
    \end{aligned}
\end{equation}

Though not entirely obvious, this is actually   de Sitter space in (rotating) oblate spheroidal coordinates. 

In this $m\to0$ limit, it is easy to check that the Weyl tensor is zero (and, therefore, the Weyl scalar will be zero too). The Kretschmann scalar is found to be 
\begin{equation}
    \riem{\mu\nu\alpha\beta}{}\riem{}{\mu\nu\alpha\beta} = \frac{8}{3}\,\Lambda^2\,.
\end{equation}
Note, that (as expected) this is (trivially) the \textit{difference}  of the Kretschmann scalar and Weyl scalar, as for the KdS and Kottler cases.

One may now go one step further and perform an explicit  coordinate transformation on the line element \eqref{m=0 form} to obtain the ``standard" form of the de Sitter metric. Using the explicit coordinate transformation given in reference~\cite{Matzner}
\begin{equation}
    \begin{aligned}
        & T = \frac{t}{\Xi};\\
        & \Phi = \phi - \frac{a\Lambda t}{3\Xi};\\
        & y\cos\Theta = r\chi;\\
        & y^2  = \frac{1}{\Xi}\Big[r^2\metcomp{\theta} + a^2(1-\chi^2)\Big]\,,\\
    \end{aligned}
\end{equation}
one can show that \eqref{m=0 form} reduces to
\begin{equation}\label{reduced kds}
    \ds{\mu}{\nu} = -(1-\frac{\Lambda}{3}y^2)\thin \D T^2 + \frac{1}{1-\frac{\Lambda}{3}y^2}\thin \D y^2 + y^2 \D \Theta^2 + y^2 \sin^2 \Theta\, \D\Phi^2\,. 
\end{equation}
This is in fact the standard form for de Sitter space, presented in terms of the coordinates $(T,y,\Theta,\Phi)$, which we could simply re-name $(t,r,\theta,\phi)$ if desired.

Performing two further coordinate transformations allows us to cast this metric into a form where --- explicitly --- space is exponentially expanding. First, we transform the time coordinate according to 
\begin{equation}
    T = \Tilde{t} + \int \frac{Hy}{1-H^2 y^2} \D y = \Tilde{t} + \frac{\ln{(1-H^2y^2)}}{2H}\,,
\end{equation}
resulting in the Painlev\'e--Gullstrand \cite{PG-cosmology} form of de Sitter space:
\begin{equation}
    \ds{\mu}{\nu} = -\D \Tilde{t}\,^2 + [\,\D y - Hy\,\D t \,]^2 + y^2 \D \Omega^2\,.
\end{equation}
Secondly, we transform the radial coordinate according to 
\begin{equation}
    y = \Tilde{r}e^{Ht}\,,
\end{equation}
resulting in de Sitter space in comoving coordinates: 
\begin{equation}
    \ds{\mu}{\nu} = -\D \Tilde{t}\,^2 + e^{\,2H\Tilde{t}}\{\thin\D \Tilde{r}^2 + \Tilde{r}^2 \D \Omega^2\thin\}\,.
\end{equation}
Therefore, it is apparent that, as desired, the $m\to0$ limit of Kerr--de Sitter is indeed the exponentially growing FLRW spacetime. 
We again see that the only even slightly tricky part of the analysis was in setting up the coordinate transformations used to make these properties manifest. 

%====================================================
\section{Asymptotic behaviour of Kerr-de Sitter}
%====================================================
The claim that the mass of a black hole grows as a function of time has been proven to be false thus far. In sub-section \ref{SS:Kottler} we have shown that for the Schwarzschild--de~Sitter (Kottler) spacetime the mass of the central black hole is simply $m$, a time independent constant. 
In sub-section \ref{SS:McVittie} we obtained the same result for McVittie (Schwarzschild--FLRW) spacetime. 
We shall now show that this also holds for the Kerr--de~Sitter black hole  by considering the asymptotic behaviour of the Kerr--de~Sitter spacetime. 

%=====================================
\subsection{Small $r$ expansion}\enlargethispage{40pt}
%=====================================

Let us begin with `small' $r$, i.e., $|\Lambda| r^2\ll 1 $, while keeping  $ r>a$. It is perhaps obvious that we should expect --- when close to the central black hole --- the metric to be of the form ``Kerr $+$ perturbation". For the following analysis we shall use the binomial expansion in $r$. We note:
\begin{equation}    
    \begin{aligned}
        &\frac{1}{\Xi^2} \approx 1 - \frac{2}{3}\Lambda a^2\,;\\
        & \frac{1}{\metcomp{\theta}} \approx 1 - \frac{1}{3}\Lambda a^2 \chi^2\,.
    \end{aligned}
\end{equation}
Component by component we explicitly find 
\begin{equation}
    \begin{aligned}
         &(g_{tt})_{KdS} \approx (g_{tt})_{Kerr} - \frac{2}{3}\Lambda a^2\,(g_{tt})_{Kerr} + \laund{\Lambda a^2 } = (g_{tt})_{Kerr} \Big( 1 + \laund{\Lambda a^2 }\Big);\\
         &(g_{rr})_{KdS} \approx (g_{rr})_{Kerr} + \frac{1}{3}\Lambda r^2\,(g_{rr})_{Kerr} + \laund{\Lambda r^2 } = (g_{rr})_{Kerr} \Big( 1 + \laund{\Lambda r^2 }\Big);\\
         &(g_{\theta\theta})_{KdS} \approx (g_{\theta\theta})_{Kerr} - \frac{1}{3}\Lambda a^2\thin \chi^2 \,(g_{\theta\theta})_{Kerr} + \laund{\Lambda a^2 \chi^2 } = (g_{\theta\theta})_{Kerr} \Big( 1 + \laund{\Lambda a^2 \chi^2 }\Big);\\
         &(g_{\phi\phi})_{KdS} \approx (g_{\phi\phi})_{Kerr} - \frac{2}{3}\Lambda a^2 \,(g_{\phi\phi})_{Kerr} + \laund{\Lambda r^2 } = (g_{\phi\phi})_{Kerr} \Big( 1 + \laund{\Lambda r^2 }\Big);\\
         &(g_{\phi\thin t})_{KdS} \approx (g_{\phi\thin t})_{Kerr} - \frac{2}{3}\Lambda a^2 \,(g_{\phi\thin t})_{Kerr} + \laund{\Lambda r^2 } = (g_{\phi\thin t})_{Kerr} \Big( 1 + \laund{\Lambda r^2 }\Big)\,.
    \end{aligned}
\end{equation}
Note that the $g_{rr}$ term is not merely a straightforward binomial expansion in $r$. Rather, we use the fact that in the region of interest
\begin{equation}
    (g_{rr})_{KdS} = \frac{\rho^2}{(r^2 +a^2)(1-H^2 r^2) - 2mr} \approx \frac{\rho^2}{r^2 + a^2 - 2mr}\; \frac{1}{1- H^2 r^2}\,,
\end{equation}
which is true as  we can safely neglect $\laund{r^3}$ terms. 
Finally, since $\chi \in [-1,1]$ and we have assumed $a<r$, all the indivudual components of the Kerr--de Sitter metric may be written as
\begin{equation}
    (g_{\mu\nu})_{KdS} = (g_{\mu\nu})_{Kerr} \Big( 1 + \laund{\Lambda r^2 }\Big)\,.
\end{equation}
Consequently at small distances (meaning $|\Lambda| r^2 \ll 1$)  Kerr--de~Sitter reduces to Kerr as expected --- with a constant unchanging mass parameter $m$, and no sign of any direct coupling between the de Sitter expansion and the central black hole.  

%====================================
\subsection{Large $r$ expansion}
%====================================

For the large $r$ expansion we assume $r\gg m$ (we also assume $m>a$ to avoid naked singularities). 
However we do not want $r$ to become cosmologically enormous, we still want to keep $|\Lambda| r^2 \lesssim 1$. 
(If $\Lambda>0$ one certainly does not want to go past the cosmological horizon at $r_C\approx1/ \sqrt{\Lambda}$. 
In counterpoint, if $\Lambda<0$, corresponding to an asymptotically anti-de Sitter space, there is simply no need to go past $r \sim 1/\sqrt{|\Lambda|}$ to detect cosmological physics.)

As we are a suitably large (but not too large) distance away from the black hole, one would expect the metric to be of the form ``de Sitter + perturbation". For all of the metric components except the $g_{rr}$ component, we may easily separate out the mass terms and then expand about large $r$:
\begin{equation}
    \begin{aligned}
        &(g_{tt})_{KdS} = (g_{tt})_{dSO} + \frac{2mr}{\rho^2\Xi^2} \approx (g_{tt})_{dSO} +\frac{2m}{r} + \laund{\frac{m}{r^3}};\\
        &(g_{\theta\theta})_{KdS} = (g_{\theta\theta})_{dSO} \approx (g_{\theta\theta})_{dSO}  + \laund{\frac{2m}{r}};\\
        &(g_{\phi\phi})_{KdS} = (g_{\phi\phi})_{dSO} + \frac{2mra^2(1-\chi^2)^4}{\rho^2\Xi^2}\frac{1}{\Xi^2} \approx (g_{\phi\phi})_{dSO} + \frac{2m}{r}\frac{1}{\Xi^2} a^2(1-\chi^2)^4 + \laund{\frac{2m}{r^3}};\\
        &(g_{\phi\thin t})_{KdS} = (g_{\phi\thin t})_{dSO} + \frac{2mra(1-\chi^2)^2}{\rho^2\Xi^2}\frac{1}{\Xi^2} \approx (g_{\phi\thin t})_{dSO} + \frac{2m}{r}\frac{1}{\Xi^2} a(1-\chi^2)^2 + \laund{\frac{2m}{r^3}}\,.\\ 
    \end{aligned}
\end{equation}
Here, the subscript $dSO$ is used to make it explicit that this is a component of the de~Sitter metric \emph{in oblate spheroidal coordinates}. The $(g_{rr})_{KdS}$ component is most easily dealt with by writing:
%using a (physically equivalent) expansion about $m=0$ instead of large $r$. Performing the expansion we find 
\begin{eqnarray}
    (g_{rr})_{KdS} &=& (g_{rr})_{dSO} - \frac{2mr\thin(r^2 + a^2\chi^2)}{(r^2 + a^2)^2\thin(1-H^2 r^2)^2} 
    \nonumber\\
    &\approx& (g_{rr})_{dSO} - \frac{2m}{r}\frac{1}{(1-H^{2} r^{2})^2} \left[ 1 + \laund{\frac{a^2}{r^2}}\right]\,.
\end{eqnarray}
Therefore, we may write the Kerr--de Sitter metric expanded about large $r$ ($r\gg m$, but $r$ not cosmologically large, 
$|\Lambda| r^2 \lesssim 1$) as
\begin{equation}
    (g_{\mu\nu})_{KdS} = (g_{\mu\nu})_{dSO} + \laund{\frac{2m}{r}}\,.
\end{equation}

As required, as we approach large (but not too large) $r$ the Kerr--de~Sitter spacetime asymptotically approaches  de~Sitter space. (Which, as we have already seen, after suitable coordinate transformations can be recast in terms of an exponentially growing scale factor $a(t)=\exp(Ht)$.) 

We emphasize (again) that in this specific Kerr--de~Sitter example there simply is no coupling between the mass parameter $m$ and the cosmological parameter $H$; they are independent constants.  
The only even slightly tricky part of the analysis was in setting up the coordinate transformations used to make these properties manifest. 

%\newpage
%====================================
\section{Kerr--FLRW spacetime?}\label{Kerr--FLRW_section}
%====================================

While we have seen that Kerr--de~Sitter, corresponding to specifically exponential expansion at asymptotic spatial infinity, can be written down explicitly in a not too complicated form, we know of no equivalent result for Kerr--FLRW for a general scale factor ${a}(t)$. 
There is a reason for this: in Kerr's original article~\cite{Kerr:1963} he asked whether it would be possible to find a (perfect fluid) interior solution for what is now called Kerr spacetime. This is a question that still remains open after 60 years. Only partial results are known, in terms of anisotropic non-perfect fluids and other anisotropic sources~\cite{Israel:1968,Israel:1970,Majidi:2017}. 
Finding an exact Kerr--FLRW spacetime would be tantamount to finding a time-dependent perfect fluid exterior solution to the Kerr black hole --- which would be at least as hard as the still unsolved problem of finding a perfect fluid interior solution.

However, as mentioned in section \ref{scale_section}, the largest known galactic black holes have masses of order  $3\times 10^{10}\; m_\odot$. This corresponds to a Schwarzschild radius $\lesssim 10^{-3}$ parsec, whereas the statistical scale of homogeneity is of order $\gtrsim 10^{8}$ parsecs. Therefore, having a solution that \textit{asymptotes} to a perfect fluid on scales such that the FLRW solution is applicable is certainly \textit{good enough}. 

Furthermore, observational evidence strongly suggests that the universe is currently cosmological constant dominated, so the relevant FLRW spactime, now and for the foreseeable future, is de Sitter. Thence the Kerr-de~Sitter solution is, for all practical purposes, certainly \textit{good enough}. 

%====================================
\section{Black hole internal structure?}\enlargethispage{40pt}
%====================================

As part of the plausibility argument for entertaining a possible direct black-hole/ cosmology coupling,  reference~\cite{Farrah:2023} suggested that this might have something to do with an assumed non-trivial internal structure for black holes. Specifically, was dark energy \emph{inside} the black hole slowly being released? Several authors have tried to make this idea more precise. Whole certainly there is widespread agreement that \emph{regular black holes} and more generally black holes with a non-vacuum interior are of interest~\cite{Mottola:2023,Mazur:2001,Mazur:2004, Visser:2003, Cattoen:2005,  Carballo-Rubio:2023,Carballo-Rubio:2022,Carballo-Rubio:2018, Simpson:2019}, there is much less agreement as to whether such black hole variants directly couple to the cosmology they are embedded in. Most investigations suggest there is no such direct coupling~\cite{Cadoni:2023,Chu:2023,Casadio:2023,Gao:2023}, and the few investigations that suggest there is such an effect yield predictions that are quantitatively and qualitatively at variance~\cite{Lopez:2023} with the original proposal of reference~\cite{Farrah:2023}. 

\clearpage
%==========================
\section{Conclusions}
%==========================
What have we learnt from this discussion? Starting from three relatively well-known exact solutions to the Einstein equations (Kottler, McVittie, Kerr-de~Sitter) all of which successfully embed black holes in suitable  FLRW background, we have seen that these exact solutions exhibit no evidence of any ``direct  coupliing'' between the black hole mass and the cosmological expansion. 
Furthermore, several purely phenomenological investigations have similarly failed to find evidence for any ``direct  coupliing'' between the black hole mass and the cosmological expansion. 
Indeed the enormous separation of scales between milli-parsec black hole physics and giga-parsec cosmological physics renders any such ``direct coupling'' (\emph{independently of accretion or mergers}) grossly implausible.
We, therefore, urge extreme caution and care when mooting such ideas. 

\bigskip
\hrule\hrule\hrule

\addtocontents{toc}{\bigskip\hrule}

%=====================================================
\section*{Acknowledgements}
%=====================================================

RG was supported by a Victoria University of Wellington PhD Doctoral Scholarship.
\\
MV was directly supported by the Marsden Fund, 
via a grant administered by the Royal Society of New Zealand.

\clearpage
%=====================================================

\null
\vspace{-50pt}
\setcounter{secnumdepth}{0}
\section[\hspace{14pt}  References]{}
%
%===================================================== 

%==================================================================
\end{document}